\documentclass[11pt]{article}

\usepackage[T1]{fontenc}
\usepackage[utf8]{inputenc}
\usepackage{lmodern}
\usepackage[a4paper,margin=2.5cm]{geometry}
\usepackage{microtype}
\usepackage{amsmath,amssymb}
\usepackage{booktabs}
\usepackage{tabularx}
\usepackage{longtable}
\usepackage{array}
\usepackage{makecell}
\usepackage{graphicx}
\usepackage{xcolor}
\usepackage{enumitem}
\usepackage[numbers,sort&compress]{natbib}
\usepackage{xurl}
\usepackage[colorlinks=true,allcolors=blue!55!black]{hyperref}
\usepackage[nameinlink,noabbrev]{cleveref}
\usepackage{caption}
\usepackage{placeins}
\usepackage{orcidlink}

\providecommand{\doi}[1]{\url{https://doi.org/#1}}
\newcommand{\ManuscriptPDFAuthors}{}

\hypersetup{
  pdftitle={Loading history and window geometry bound compact-state slip ranking during granular shear startup},
  pdfauthor={\ManuscriptPDFAuthors},
  pdfsubject={Temporal alignment of granular-slip ranking under loading-history and window-geometry controls},
  pdfkeywords={granular slip, loading history, window geometry, non-affine motion, discrete-element method, average precision}
}

\setlist{nosep,leftmargin=*}
\newcommand{\Dmin}{D^2_{\min}}
\newcommand{\keywords}[1]{\par\noindent\textbf{Keywords:} #1\par}

\title{Loading history and window geometry bound compact-state slip ranking
during granular shear startup}
\author{Ruixin Zhou$^{1}$\,\orcidlink{0009-0004-9303-6674}
and Boliang Yu$^{1}$\thanks{Corresponding author. Email: \texttt{yuboliang@sjtu.edu.cn}}\,\,\orcidlink{0009-0006-4664-5068}\\[0.6em]
\small $^{1}$School of Physics and Astronomy, Shanghai Jiao Tong University, Shanghai 200240, China
}
\date{}

\begin{document}

\maketitle

\begin{abstract}
Granular slip forecasting can conflate material state, loading progress, and the geometry of
event-centered sampling. We separated these contributions in slowly sheared two-dimensional
frictional disks using a compact neural score of stress, pressure, coordination, non-affine
motion, and force-network observables. The model was developed on 36 trajectories and frozen
before testing on 18 new trajectories under two nested stress-drop definitions. Inspection
of held-out results revealed post-event sampling asymmetry; recovery-aware analyses are
therefore descriptive. With trajectories weighted equally, the compact score ranked near-
slip windows above both prevalence and within-trajectory circular-phase controls under
both definitions (representative average precision 0.310 versus prevalence 0.173; phase-null
upper bound 0.257). Loading-history coordinates ranked more strongly, reaching 0.534
for causal elapsed strain. The recovery-aware rule retained 78.4\% of activity-gated events
and preferentially selected longer preceding intervals; ranking by time since the previous
catalogued event remained compatible with a count-conditioned geometry null. Compact
observables thus contain temporally aligned slip information, but stronger loading-history
baselines and window-geometry sensitivity bound that evidence. These startup data do not
isolate a state-specific short-horizon precursor beyond loading history or support a renewal
interpretation of elapsed-strain ranking.
\end{abstract}

\section{Introduction}
\label{sec:introduction}

Dense granular materials store load in heterogeneous contact networks and
release it through abrupt irreversible rearrangements. Slowly driven systems
show large stress fluctuations, broadly distributed avalanches, and plastic
events ranging from local transformations to system-scale organization
\citep{howell1999stress,dahmen2011avalanches,maloney2006athermal}. Predicting
such releases poses two distinct questions. A score may attain average
precision (AP) above its no-skill positive-window prevalence, yet provide no
information beyond where the system lies in its loading protocol.
Demonstrating a material-state precursor requires the stronger comparison with
loading-history variables at the same observation times.

Particle motion and force-network structure provide plausible state
observables. The non-affine measure $D^2_{\min}$ detects departures from local
affine deformation \citep{falk1998dynamics}, while photoelastic and numerical
studies connect stick--slip to force-network change and microslip activity
\citep{daniels2008force,picaciamarra2010unjamming,ferdowsi2013microslips}.
Network topology retains spatial organization lost by bulk averages
\citep{bassett2015extraction,papadopoulos2016evolution,
giusti2016topological}. Machine-learning studies have inferred frictional
state, time to failure, or local rearrangement propensity from acoustic and
particle-resolved signals
\citep{rouetleduc2017machine,ren2019machine,cubuk2015identifying,
bapst2020unveiling}. Time-dependent network measures have also changed before
slip in boundary-driven and intruder geometries
\citep{kramar2022intermittency,bretz2023stochastic,basak2023force}. These
results motivate state-based ranking, but do not establish that state
descriptors improve the timing of a system-level release over loading history
in homogeneous strain-controlled shear.

Loading history is a demanding baseline during startup. Cumulative strain
tracks stress buildup, fabric evolution, and the evolving slip propensity. The
strain elapsed since a previous release provides a recurrence coordinate.
Cumulative strain requires no particle measurement. The elapsed-strain score
does not use the compact state vector at the observation endpoint or any future
sample, but it does require past observations sufficient to recognize the
preceding release. Under the activity-gated definition, these include past
non-affine activity. We enforce causal availability by replaying the fixed
event detector on each observed trajectory prefix. A strong elapsed-strain
rank is not by itself evidence for a renewal mechanism: the forecast horizon,
event exclusion, and right censoring can produce catalog-age ranking even when
event onsets are randomized.

A second problem concerns which windows enter the comparison. In the original
rule, positive windows could appear sooner after a slip than negative windows.
A score can then look skillful because of the sampling asymmetry rather than a
physical precursor. A correction can also change which events contribute any
positive window and thereby select a different inter-event distribution.
Pooling all windows further mixes two effects: ranking time within one
trajectory and differences in positive-window fraction between trajectories.
We therefore report pooled performance separately from mean per-trajectory
AP, which gives every independently prepared trajectory equal weight, and
audit event coverage separately from window counts.

We study two-dimensional bidisperse frictional disks under homogeneous simple
shear at a fixed strain rate with Lees--Edwards periodic boundaries. A
six-observable compact model and a thresholded-contact graph model were
developed using a 36-trajectory development cohort. Their architectures,
parameters, normalizers, and evaluation protocol were fixed before the
18-trajectory, condition-balanced cohort was generated; no model was retrained.
Two nested stress-drop definitions test whether direct use of non-affine
activity in event acceptance controls the result. The original
prespecified analysis pooled admissible windows. After the held-out outcomes
were inspected, a post-test audit identified the post-event sampling
asymmetry. We retain the original result to document the analysis chronology.
We then applied the same previous-event exclusion to both window types. A
stricter lookback-aware rule also places every frame used by $D^2_{\min}$ after
event recovery. In both cases, AP is averaged over trajectories and
uncertainty is calculated by exact condition-stratified resampling.

The compact model ranks near-slip windows above the mean positive-window
fraction under both event definitions and above a within-trajectory
circular-phase randomization on the lookback-aware windows. Thus its signal is
temporally aligned with the held-out targets rather than being an arbitrary
phase of the same score sequence. Cumulative strain and causal prefix-replay
elapsed strain nevertheless rank the windows more accurately. Moreover, the
lookback-aware subset preferentially retains events with longer preceding
intervals, and the magnitude of final-catalog age ranking lies inside a
count-conditioned window-geometry null. These analyses were specified after
the held-out outcomes were inspected and are descriptive sensitivities, not
new confirmation. Together they establish a bounded result: the tested state
sequence contains short-horizon ranking information, but the present startup
protocol does not isolate an incremental material-state precursor beyond
loading history. Exploratory spatial analyses of non-affine motion during
release are reported separately in Appendix~\ref{app:exploratory-audits}; they
are not used to support the forecasting claim.
\section{Methods}
\label{sec:methods}

\subsection{Granular system and shear protocol}
\label{sec:dem}

We simulated $N=512$ frictional disks in two dimensions. Equal numbers of
particles had radii $r_s=0.5$ and $r_l=0.7$; the areal material density was
one, giving mass $m_i=\rho\pi r_i^2$ and moment of inertia
$I_i=m_i r_i^2/2$. Contacts followed a linear Cundall--Strack law
\citep{cundall1979discrete}. For overlap $\delta$, normal relative velocity
$v_n$, and tangential spring displacement $\delta_t$,
\begin{align}
 f_n&=\max(k_n\delta-c_n m_{\mathrm{eff}}v_n,0),\\
 f_t&=-k_t\delta_t-c_t m_{\mathrm{eff}}v_t,
 \qquad |f_t|\leq\mu f_n .
 \label{eq:contact-law}
\end{align}
Particle rotation contributed to $v_t$. Translational and rotational drags
were $-\zeta m_i\mathbf v_i$ and $-\zeta_rI_i\omega_i$. The parameters were
$k_n=5000$, $k_t/k_n=0.5$, $c_n=40$, $c_t=20$, and
$\zeta=\zeta_r=2$. Numerical-regime and stored-contact checks are reported in
Appendix~\ref{app:protocol}.

Lees--Edwards boundaries imposed homogeneous simple shear with peculiar
velocities \citep{lees1972computer}. Configurations were prepared from
$\phi=0.58$, compressed in 16 stages, and consolidated until kinetic-energy,
force-imbalance, and pressure-stability criteria were met. Production used
$\dot\gamma=3\times10^{-3}$ to total strain 0.50 and an integration step
$4\times10^{-4}$. States were saved every
$\Delta\gamma\simeq0.001$, giving 500 frames per trajectory. The stored stress
was
\begin{equation}
 \boldsymbol\sigma=\frac{1}{L_xL_y}\sum_{i<j}
 \mathbf r_{ij}\otimes\mathbf f_{ij}
 +\frac{1}{L_xL_y}\sum_i m_i\mathbf v_i\otimes\mathbf v_i,
 \label{eq:stress-tensor}
\end{equation}
with shear component
$\sigma_{xy}=(\sigma_{xy}^{\mathrm{tensor}}+
\sigma_{yx}^{\mathrm{tensor}})/2$ and pressure
$P=(\sigma_{xx}+\sigma_{yy})/2$.

The design crossed $\phi\in\{0.84,0.85,0.86\}$ with
$\mu\in\{0.1,0.3,0.5\}$. Four preparations per condition formed a
36-trajectory development cohort; two new preparations per condition formed
the 18-trajectory held-out cohort. All 54 trajectories passed the numerical
quality criteria. Models and normalization constants were fitted only in the
development cohort.

\subsection{Non-affine motion and slip definitions}

For particle $i$, non-affine motion over backward lag $\Delta t$ was computed
in the Falk--Langer form \citep{falk1998dynamics},
\begin{equation}
 D_i^2(t,\Delta t)=\frac{1}{N_i}\min_{\mathbf J_i}
 \sum_{j\in\mathcal N_i}
 \left\|\mathbf r_{ij}(t)-\mathbf J_i\mathbf r_{ij}(t-\Delta t)\right\|^2 .
 \label{eq:d2min}
\end{equation}
Reference neighbors lay within 2.2 small-particle diameters and the lag was
$\Delta\gamma=0.010$. We use $D_i^2$ for the particle value,
$\langle D_i^2\rangle_i$ for the particle mean, and the mean over the most
active 10\% of particles for the event-activity trace.

Yield was the first complete prominent stress peak after
$\gamma=0.005$ satisfying the fixed height and prominence criteria. Slip
search began strictly after yield. A candidate onset was a local peak before a
complete stress drop and recovery. The activity-gated definition required:
(i) pressure-normalized drop at least 0.010, (ii) normalized peak prominence at
least 0.005, (iii) recovery of at least 10\% within two frames, and
(iv) event-interval activity at least 15\% above its five-frame pre-event
median. Candidates separated by less than $\Delta\gamma=0.010$ were merged by
retaining the larger drop. The activity-ungated definition removed only
criterion (iv). Both definitions, including smoothing and tail censoring,
were fixed before held-out trajectories were generated. The activity-gated
catalog contained 88 onsets; the activity-ungated
catalog contained 106 and retained every gated
onset at the same frame. These final catalogs are retrospective because event
acceptance uses centered smoothing, a finite drop-search interval, recovery,
and candidate merging. Section~\ref{sec:representations} defines a separate
prefix-replay score that never reads beyond the observation time.

\subsection{Forecast target and window selection}
\label{sec:samples}

For a four-frame input window ending at frame $t$, which we call its
observation endpoint, the target was
\begin{equation}
 y_t=\mathbf 1\!\left(0<\gamma_s-\gamma_t\leq0.015
 \text{ for the next accepted slip }s\right).
 \label{eq:forecast-target}
\end{equation}
The original window set, fixed before independent testing and hereafter called
the frozen set, omitted histories containing an onset or any of its five
immediate post-event frames. A negative window additionally had
to place its complete input more than $\Delta\gamma=0.020$ from every accepted
onset; near-event ambiguity windows and right-censored tails were excluded.
Positive windows did not receive the same previous-event separation. With a
four-frame history, positives could therefore re-enter nine saved frames after
an onset, whereas the first observed negative window occurred at offset 25.

We retain the original windows to report the analysis chronology and define
two post-test subsets. Let $H=4$, $t_0=t-H+1$, and let $s^{-}(t)$ be the latest
onset no later than $t$. The median saved strain increment was
$9.996\times10^{-4}$ in every held-out trajectory. The fixed
$\Delta\gamma=0.020$ buffer therefore corresponded to $b_j=21$ saved-frame
intervals. The common-exclusion rule requires
\begin{equation}
 e_t^{\mathrm{sym}}=1
 \quad\Longleftrightarrow\quad
 s^{-}(t)\text{ is absent, or }
 t_0-s^{-}(t)>b_j .
 \label{eq:symmetric-eligibility}
\end{equation}
This requirement is applied identically to positive and negative windows. The
second rule also accounts for the backward look used to compute non-affine
motion. Let $r_{D^2}(k)$ be the stored reference-frame index used by $D_i^2$ at
input frame $k$, and let $u^{-}(t)$ be the trough of the previous event. The
lookback-aware rule additionally requires
\begin{equation}
 \min_{k=t_0,\ldots,t}r_{D^2}(k)>u^{-}(t)+2 .
 \label{eq:support-eligibility}
\end{equation}
The two-frame term covers the fixed recovery decision. The lookback-aware rule
is the intersection of Eqs.~\eqref{eq:symmetric-eligibility} and
\eqref{eq:support-eligibility}. It can remove but never add a frozen window.
No frozen label, event target, window, or prediction is recomputed.

\begin{figure*}[tb]
  \centering
  \includegraphics[width=\textwidth]{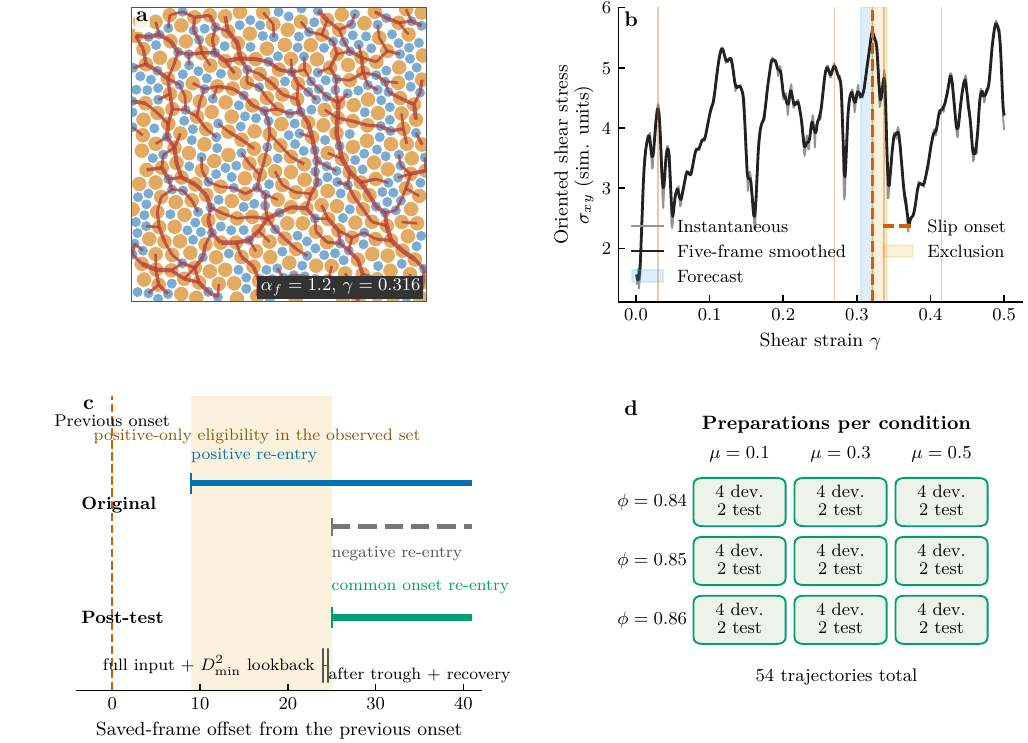}
  \caption{Simulation and evaluation design. (a)~Bidisperse configuration;
  displayed edges satisfy $f_n\geq1.2\langle f_n\rangle$ and
  periodic-crossing edges are omitted from the drawing only.
  (b)~Instantaneous and five-frame-smoothed stress, one accepted onset, its
  forecast interval, and the original event-exclusion interval.
  (c)~Schematic of the observed
  original post-event asymmetry for $H=4$. Applying the original negative-window
  buffer to positive and negative windows gives a common onset-based re-entry
  at offset 25; the
  lookback-aware rule can move it later until the complete $D^2_{\min}$
  lookback follows trough and recovery. (d)~Each
  $(\phi,\mu)$ cell contributes four development and two held-out
  trajectories}
  \label{fig:system-protocol}
\end{figure*}

Figure~\ref{fig:system-protocol} summarizes the system, event window, observed
sampling asymmetry, post-test rules, and condition-balanced cohorts.

\subsection{State models and loading-history scores}
\label{sec:representations}

The compact input contained four frames of six scalar observables,
\begin{equation}
 \mathbf x_{\mathrm{compact}}=
 (\sigma_{xy},P,Z,\langle D_i^2\rangle_i,f_{\max}^{n},S_{\max}),
 \label{eq:compact-features}
\end{equation}
where $Z$ is mean coordination, $f_{\max}^{n}$ is maximum normal contact
force, and $S_{\max}$ is the largest component of the strong-contact graph.
The compact classifier flattened the $H=4$ input and used two 64-unit ReLU
layers. Three fixed-seed probabilities were averaged.

A secondary $H=8$ graph pipeline retained contacts with
$f_n\geq\alpha_f\langle f_n\rangle$, $\alpha_f=1.2$, encoded particle and
edge features with two mean-aggregation layers, globally mean-pooled each
frame, and passed the sequence through a 64-unit LSTM
\citep{hamilton2017inductive,hochreiter1997long}. It received no explicit
stress or pressure and changed both representation and classifier; it is not a
strict feature ablation. A stress--pressure--ratio MLP provided another
secondary comparison. Architectures, normalization values, parameter counts,
and training histories are given in Appendix~\ref{app:protocol}.

The principal loading-history scores required no fitting. Cumulative strain
used $\gamma_t$. To construct elapsed strain without future observations, we
replayed the complete fixed label algorithm separately on every trajectory
prefix $\{0,\ldots,t\}$. If that replay returned an onset, $s_{\rm p}(t)$ was
its latest onset and the score was
\begin{equation}
 e_t^{\rm p}=\gamma_t-\gamma_{s_{\rm p}(t)} .
 \label{eq:causal-elapsed}
\end{equation}
The score was zero before the first prefix-detected event. Replaying a complete
trajectory reproduced its stored yield, onset, and trough arrays exactly under
both event definitions. The prefix replay was introduced after outcome
inspection and is therefore a causal-availability sensitivity, not a new
prespecified baseline. All scores were evaluated on identical frozen window
keys within an event definition and window rule.

\subsection{Evaluation summaries and uncertainty}
\label{sec:metrics}

Average precision (AP) was used because positive windows were a minority
\citep{saito2015precision}. The original analysis, fixed before testing,
pooled every admissible held-out window. It asked whether compact-model AP
exceeded the pooled positive-window fraction for each event definition. We
retain those estimates to document the analysis chronology. When trajectories
have different positive-window fractions, however, pooled AP mixes temporal
ranking within a trajectory with ranking between trajectories.

The post-test analysis instead computes AP separately for every trajectory and
then gives the 18 trajectories equal weight. We call this mean per-trajectory
AP and denote it by $\mathrm{AP}_{\mathrm{macro}}$:
\begin{equation}
 \mathrm{AP}_{\mathrm{macro}}=\frac{1}{18}\sum_{j=1}^{18}\mathrm{AP}_j,
 \qquad
 \pi_{\mathrm{macro}}=\frac{1}{18}\sum_{j=1}^{18}\pi_j .
 \label{eq:macro-ap}
\end{equation}
Under every reported window rule and event definition, all 18 trajectories
retained at least one positive and one negative window, so every
$\mathrm{AP}_j$ was defined.
Compact-minus-prevalence, compact-minus-cumulative-strain, and
compact-minus-elapsed-strain differences were first formed within trajectory.
Uncertainty uses exact condition-stratified resampling. We draw two
trajectories with replacement in each of the nine $(\phi,\mu)$ cells. Only
$3^9=19,683$ distinct compositions exist, so all are
evaluated with their exact probabilities. This removes Monte Carlo noise from
the 95\% intervals, but does not create additional independent preparations or
population-level confirmation.
The intervals condition on the frozen three-seed ensemble probabilities and do
not propagate model-training-seed variation.

For a secondary interpretation check, we computed the Spearman correlation
between compact probability and cumulative strain separately in each
trajectory, then averaged the 18 correlations with equal trajectory weight.
Its interval uses the same exact resampling design.

The window-selection, trajectory-balanced, prefix-replay, and temporal-null
analyses were specified after the held-out results were inspected. They are
descriptive sensitivities, not newly prespecified confirmation, and their
intervals are not adjusted for multiplicity. A trajectory-constant
observed-prevalence score is used only to diagnose the susceptibility of pooled
AP to differences in trajectory prevalence; because it uses evaluation labels,
it is not a predictor. The frozen protocol, per-trajectory results,
machine-readable window masks, and prefix-replay scores are included in the
archived data and code record.

\subsection{Post-test event-selection and temporal-null audit}
\label{sec:temporal-null-methods}

We counted an event once under a window rule if at least one retained positive
endpoint targeted its onset. Preceding-onset intervals were summarized with
each represented event weighted once; first events were excluded because they
have no preceding onset. This distinguishes a change in the represented event
population from the number of overlapping positive windows.

For the compact-score temporal control, lookback-aware endpoints were ordered
by observation frame within each trajectory. We computed AP after every
possible integer circular shift of that trajectory's unchanged compact-score
sequence. Each of 100,000 null cohorts selected one shift
independently and uniformly for each trajectory and averaged the resulting 18
AP values. The central 95\% randomization range tests alignment against this
specific phase control; it neither removes all loading-stage information nor
replaces the mean positive-window fraction as the conventional random-ranking
reference.

A separate geometry calibration generated 2,000
count-conditioned surrogate cohorts. Within each trajectory it retained the
observed event count, event-duration multiset, yield and right-censor frames,
strain grid, and non-affine validity/reference arrays. Onsets were sampled
uniformly over the detector-supported post-yield frames subject to the fixed
$\Delta\gamma=0.010$ merge separation, and the observed durations were
randomly permuted. We then rebuilt all $H=4$ endpoints and applied the original
forecast horizon, event buffers, complete-$D^2_{\min}$ requirement, common
exclusion, and lookback-after-recovery rule. The ideal score was catalog age
since the latest surrogate event whose trough plus recovery was observable.
This procedure calibrates final-catalog window geometry; it is not a surrogate
stress history, a causal prefix detector, or a goodness-of-fit test for a
Poisson or renewal process. Supplying the actual catalogs reproduced every
frozen endpoint key, target, target event, lookback mask, and catalog-age score
exactly before surrogate results were accepted.

Exploratory release-localization, alternative event-gate, post-yield, and
graph-pipeline analyses are separated from the forecasting comparison and
reported in Appendix~\ref{app:exploratory-audits}.

\section{Results}
\label{sec:results}

\subsection{The original window set contains a post-event sampling asymmetry}
\label{sec:result-eligibility}

The original held-out analysis contained 6,166
activity-gated windows, of which 1,230 were
positive, and 5,945 activity-ungated
windows, of which 1,396 were positive.
When all windows were pooled, compact-model AP was
0.284 at prevalence
0.199 and
0.305 at prevalence
0.235. These are the original
prespecified pooled estimates.

The window audit identified a one-sided post-event region. At offsets 9--24
saved frames after the previous onset, the frozen activity-gated set contained
231 windows and the activity-ungated
set contained 336; every one was
positive. They accounted for
18.8\% and
24.1\% of all positive windows.
A post-test check ranked each window only by how recently a previous accepted
onset had occurred. This recency score achieved AP
0.434 and
0.523. After removing the region, its AP fell to
0.194 and
0.227. Because the score uses the
retrospectively accepted event catalog and was constructed specifically after
outcome inspection, we treat it only as a window-geometry diagnostic, not a
prespecified or deployable forecasting baseline.

The backward non-affine lookback provided a second asymmetry. At the observation
endpoint, 32 activity-gated and
57 activity-ungated positive windows
used a $D^2_{\min}$ reference at or before the previous onset. Across the full
four-frame input, the counts rose to
81 and
131. The common-exclusion and
lookback-aware subsets were introduced to remove these one-sided channels
rather than to reinterpret them with additional caveats.

The correction also changed the represented event population. The original
positive windows targeted 88 activity-gated and
106 activity-ungated events. Lookback-aware
positive windows retained 69
(78.4\%) and 74
(69.8\%), respectively. With every retained event
weighted once and first events excluded, the median preceding interval shifted
from 48.0 to
78.0 frames for activity-gated events and
from 39.0 to
72.5 frames for activity-ungated events.
The corrected comparison therefore preferentially represents more widely
separated releases rather than the complete original event population.

\subsection{Equal-trajectory ranking is above the no-skill level but below loading history}
\label{sec:result-macro}

Tables~\ref{tab:eligibility-macro-ap} and
\ref{tab:eligibility-macro-differences} report equal-trajectory results for
the original, common-exclusion, and lookback-aware window sets. Under the
lookback-aware rule, the compact model retained
5,742 activity-gated windows and
5,439 activity-ungated windows.
Mean per-trajectory AP was
0.310 and
0.325, compared with mean
per-trajectory positive-window fractions
0.173 and
0.199. Compact minus
prevalence was 0.138
[0.083, 0.192] and
0.126
[0.080, 0.171]. The compact
score therefore retains within-trajectory ranking information after the
window correction.

Loading history ranked the same windows more accurately. Mean per-trajectory
cumulative-strain AP was 0.398 and
0.420. Compact minus cumulative
strain was -0.088
[-0.162, -0.013] and
-0.095
[-0.161, -0.029]. The causal
prefix-replay elapsed score was stronger still, with AP
0.534 and
0.508. Compact minus causal
elapsed strain was
-0.223
[-0.288, -0.159] and
-0.184
[-0.224, -0.143].

\begin{table*}[tb]
  \centering
  \caption{Mean per-trajectory AP on the held-out cohort. ``Common exclusion''
  applies the same previous-onset separation to positive and negative windows.
  ``Lookback-aware'' also requires every frame used by the four-frame
  $D^2_{\min}$ history to follow the previous trough and recovery. Elapsed AP
  uses the causal prefix replay in Eq.~\eqref{eq:causal-elapsed}.}
  \label{tab:eligibility-macro-ap}
  \footnotesize
  \setlength{\tabcolsep}{3.2pt}
  \begin{tabular}{@{}llrcccc@{}}
    \toprule
    Definition & Rule & \makecell{Windows\\(positive)} &
    \makecell{Mean positive\\fraction} & \makecell{Compact\\AP} &
    \makecell{Cumulative-\\strain AP} & \makecell{Causal\\elapsed AP} \\
    \midrule
Gated & Original & 6,166 (1,230) & 0.208 & 0.339 & 0.409 & 0.539 \\
Gated & Common exclusion & 5,935 (999) & 0.177 & 0.311 & 0.388 & 0.519 \\
Gated & Lookback-aware & 5,742 (939) & 0.173 & 0.310 & 0.398 & 0.534 \\
Ungated & Original & 5,945 (1,396) & 0.247 & 0.359 & 0.443 & 0.534 \\
Ungated & Common exclusion & 5,609 (1,060) & 0.202 & 0.326 & 0.406 & 0.499 \\
Ungated & Lookback-aware & 5,439 (1,006) & 0.199 & 0.325 & 0.420 & 0.508 \\
\bottomrule
  \end{tabular}
\end{table*}

\begin{table*}[tb]
  \centering
  \caption{Paired mean per-trajectory AP differences. Brackets are exact 95\%
  intervals from all 19,683 condition-stratified
  resampling compositions. The elapsed score is the causal prefix replay.}
  \label{tab:eligibility-macro-differences}
  \small
  \setlength{\tabcolsep}{4.5pt}
  \begin{tabular}{@{}llccc@{}}
    \toprule
    Definition & Rule & Compact $-$ prevalence & Compact $-$ strain &
    Compact $-$ elapsed \\
    \midrule
Gated & Original & 0.131 [0.076, 0.187] & -0.070 [-0.127, -0.013] & -0.200 [-0.260, -0.140] \\
Gated & Common exclusion & 0.133 [0.081, 0.186] & -0.077 [-0.143, -0.010] & -0.208 [-0.270, -0.146] \\
Gated & Lookback-aware & 0.138 [0.083, 0.192] & -0.088 [-0.162, -0.013] & -0.223 [-0.288, -0.159] \\
Ungated & Original & 0.112 [0.065, 0.158] & -0.085 [-0.133, -0.036] & -0.175 [-0.217, -0.134] \\
Ungated & Common exclusion & 0.124 [0.081, 0.167] & -0.080 [-0.138, -0.022] & -0.173 [-0.216, -0.131] \\
Ungated & Lookback-aware & 0.126 [0.080, 0.171] & -0.095 [-0.161, -0.029] & -0.184 [-0.224, -0.143] \\
\bottomrule
  \end{tabular}
\end{table*}

\begin{figure*}[tb]
  \centering
  \includegraphics[width=\textwidth]{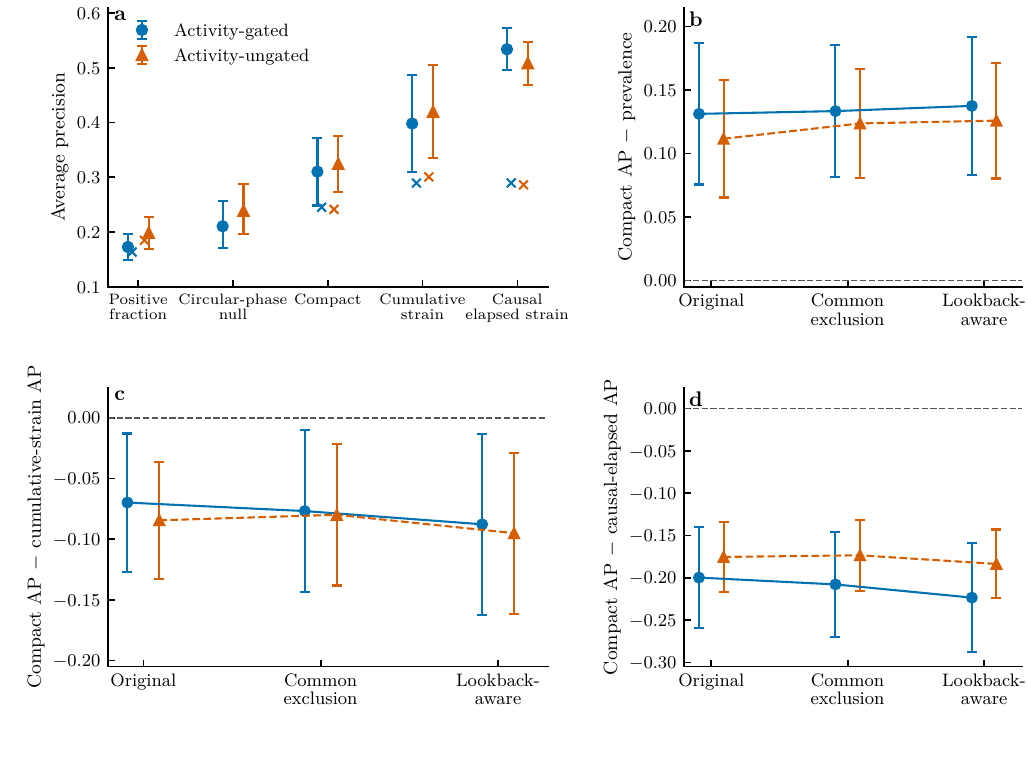}
  \caption{Loading-history comparison and window-rule sensitivity.
  (a)~Lookback-aware mean per-trajectory AP and pooled-window point estimates
  ($\times$; omitted for the phase null). Error bars are exact 95\% resampling
  intervals except for the circular-phase null, where they show the central
  95\% randomization range about its mean. (b)~Mean per-trajectory
  compact AP minus mean per-trajectory prevalence. (c)~Compact minus
  cumulative-strain AP.
  (d)~Compact minus causal prefix-replay elapsed-strain AP. Original denotes
  the initial window set; common exclusion applies the same previous-onset
  separation to both window types. The lookback-aware rule additionally places
  the complete $D^2_{\min}$ history after previous-event recovery. Every
  comparison is paired on identical windows within a trajectory}
  \label{fig:ranking-eligibility}
\end{figure*}

Figure~\ref{fig:ranking-eligibility} places the two summaries and window-rule
sensitivities in one comparison. Panel (a) also places the compact score beside
its lookback-aware circular-phase distribution. Compact minus prevalence
remains positive under every rule. Compact minus cumulative strain and compact
minus elapsed strain remain negative, and their lookback-aware intervals
exclude zero. These post-test intervals are descriptive and unadjusted for
multiplicity. The direction of the loading-history boundary is therefore
insensitive to removal of the asymmetric post-event windows.

The prefix replay used no sample after an observation endpoint. At full
trajectory length it reproduced all stored yield frames, onsets, and troughs
under both event definitions. Removing windows before the first
prefix-detected event removed
2094 activity-gated windows and
1824 activity-ungated windows. On the
remaining lookback-aware windows, causal elapsed-strain AP increased to
0.652 and
0.593. Compact-minus-elapsed
intervals remained below zero:
[-0.402, -0.261] and
[-0.298, -0.204].

\subsection{Compact alignment survives phase randomization, but catalog age is geometry-compatible}
\label{sec:result-temporal-nulls}

On the lookback-aware endpoints, the observed compact AP values were
0.310 and 0.325.
Within-trajectory circular shifting preserved each score sequence but changed
its phase relative to the target sequence. The null means were
0.211 and 0.239, with
central 95\% ranges [0.171, 0.257] and
[0.196, 0.287]. The observed-minus-null-mean differences were
0.100 and 0.086; plus-one
upper-tail probabilities were 0.00005 and
0.00051. Thus compact ranking depends on temporal
alignment beyond this structured randomization. This comparison does not
replace the positive-window fraction as the standard random-ranking baseline
and does not remove every form of loading-stage information.

The final-catalog geometry calibration gave actual catalog-age AP
0.527 and 0.504. Under the
count-conditioned hard-core onset surrogate, the corresponding null means were
0.461 and 0.503, with
central 95\% ranges [0.383, 0.538] and
[0.428, 0.572]. Both actual values lay inside those ranges.
The activity-gated value was in the upper part of its null distribution
(one-sided upper-tail probability 0.048), so the audit
does not show equivalence to the null. It does show that the observed magnitude
is not separated from the protocol-defined central geometry-null range. Because
this ideal final-catalog score differs from the causal prefix replay used in
the main comparison, the result constrains the physical interpretation of
elapsed ranking rather than replacing that comparison.

\subsection{Pooled AP also rewards differences between trajectories}
\label{sec:result-estimands}

The difference between pooled and mean per-trajectory estimates is visible in
Fig.~\ref{fig:ranking-eligibility}(a). On lookback-aware windows, pooled
compact AP was 0.245 and
0.241, whereas mean
per-trajectory AP was
0.310 and
0.325. For cumulative strain,
pooled AP and mean per-trajectory AP were
0.290 and
0.398, respectively, for the
activity-gated definition, and
0.301 and
0.420 for the activity-ungated
definition. For causal elapsed strain, the corresponding values were
0.290 and
0.534, and
0.286 and
0.508.

The ordering of the two loading-history variables depends on the summary. On the original
windows, pooled cumulative-strain AP
(0.331 and
0.364) slightly exceeded pooled
causal elapsed-strain AP
(0.329 and
0.344). Mean per-trajectory AP gave
the opposite ordering: causal elapsed strain reached
0.539 and
0.534, versus cumulative-strain AP
0.409 and
0.443. Thus elapsed strain is the
stronger within-trajectory variable, but it is not universally stronger across
summary choices.

The later strain audit on the original pooled windows gave compact minus
cumulative strain -0.047
[-0.101, 0.013] for activity-gated events and
-0.059 [-0.111, -0.004] for
activity-ungated events. The first interval includes zero. We therefore do not
attribute a general loading-history advantage to the original pooled analysis;
the stronger ordering belongs to the explicitly post-test, equal-trajectory
sensitivity.

As a check, we replaced the model score by one constant per trajectory: that
trajectory's observed positive-window fraction. This label-using score gave pooled AP
0.306 and
0.375 on the original windows. The
score has no within-trajectory timing information and uses held-out labels, so
it is not a deployable baseline. It nevertheless shows that pooled AP can be
large when trajectories differ only in prevalence. Equal-trajectory
evaluation avoids rewarding a trajectory merely for contributing more
windows. The compact probability also correlated with cumulative strain. On
the lookback-aware windows, mean within-trajectory Spearman correlations were
0.538
[0.472, 0.603] and
0.593
[0.529, 0.657]. Together with the
paired AP ordering, this supports the interpretation that the compact score
partly indexes loading history.

In a separate post-test check on the original windows, adding the fixed
compact score to a development-fitted time-plus-condition stacker changed AP
by -0.003 [-0.029, 0.011]
and -0.008
[-0.026, 0.013]. Neither interval established an
increment beyond loading time and condition; this is not an equivalence test.

\subsection{Secondary model comparisons}
\label{sec:result-secondary}

The original held-out comparisons of the stress--pressure--ratio control,
$H=8$ compact model, and thresholded graph pipeline are retained in
Appendix~\ref{app:exploratory-audits}. Neither compact--control interval
established a difference or equivalence. Thus these data provide no detected
increment from the six-channel compact model over that bulk-mechanical
control, but the comparison is too imprecise to establish equal performance.
At $H=8$, cumulative strain exceeded both compact and graph scores under both
event definitions. Because the graph omitted bulk
stress and pressure, used one contact threshold, and changed the classifier as
well as the representation, this result applies only to the tested pipeline;
it is not evidence that contact-network information is generally irrelevant.

The release-localization analysis is also confined to the Appendix. Its
duration- and selector-matched comparison covered only
68.2\% and 60.4\% of the two event
catalogs, below the fixed 80\% adequacy target, and the concentration contrast
appeared after onset. It is consequently an exploratory description of
matchable releases rather than a slip-precursor result.
\section{Discussion}
\label{sec:discussion}

The held-out trajectories support a clear but bounded result. A compact model
of bulk mechanics, coordination, non-affine activity, maximum force, and one
force-network summary attains mean per-trajectory AP above both the
corresponding mean positive-window fraction and a within-trajectory
circular-phase control. The compact sequence therefore contains information
aligned with near-slip times. It does not rank the same windows as well as
cumulative strain or causal prefix-replay elapsed strain. This ordering holds
when every trajectory is weighted equally, both window types share the same
post-event exclusion, and the non-affine lookback follows event recovery. At
the same time, that correction preferentially retains more widely separated
events, and final-catalog age ranking is compatible with the audited geometry
null. The measurable compact ranking therefore remains but is not isolated from
loading history as a state-specific precursor in the present cohort. Because
these comparisons were introduced after outcome inspection, they are
descriptive sensitivities rather than new confirmation.

\subsection{Loading history is a physical baseline during startup}

Cumulative strain is known at the observation time, but it is a loading
coordinate rather than a measured material-state channel. During startup it
tracks stress buildup, contact-network anisotropy, compaction history, and the
changing slip propensity. Elapsed strain adds a recurrence coordinate tied to
the latest release detected from observations available by the endpoint. The
full prefix replay reproduces the final catalog at trajectory completion, yet
never reads beyond an endpoint when assigning its score. Its high mean
per-trajectory AP shows that prior-event timing is strongly associated with
the fixed-window targets in the selected sample. Removing windows before the
first prefix-detected event strengthens this ordering, so their initial zero
score does not create the result. The compact score's moderate correlation
with cumulative strain is consistent with the same observables evolving along
the loading path. Adding that score to a development-fitted model of loading
time and condition produced no detected AP increment, although the intervals
do not establish equivalence.

The event catalogs contain finite inter-event scales, but the audit does not
identify their point-process origin. Their frame-interval coefficients of
variation, 0.753 and
0.837, do not test exponential waiting times
or memorylessness. More importantly, a count-conditioned homogeneous
hard-core onset surrogate reproduced a broad range of final-catalog age AP
that contained both observed values. The activity-gated value lay toward the
upper tail, so equivalence to the geometry null was not established. Conversely,
membership in the protocol-defined central range means that high catalog-age
AP alone cannot identify renewal-like modulation. The surrogate is not a
stress-history generator or a causal prefix replay; it limits the mechanism
claim rather than invalidating the observed loading-history association.

The distinction between pooled and mean per-trajectory AP matters physically.
Pooled AP describes ranking in the combined data pool. A model can therefore
gain performance by assigning generally higher scores to trajectories with
higher positive-window fractions. Mean per-trajectory AP instead asks how well
a score orders time within an equally weighted preparation. The compact
model's excess over the mean positive-window fraction shows within-trajectory
ranking. The relative order of cumulative and elapsed strain reverses between
the original pooled and equal-trajectory summaries, so neither history
variable is universally stronger. Both nevertheless exceed the compact score
in the post-test equal-trajectory analysis.

\subsection{Window selection changes the comparison}

The original target definition separated future-positive windows from
event-distant negatives, but only negatives received the full previous-event
buffer. That construction created a post-onset region containing positive
windows but no negatives in the observed cohort. Because $D_i^2$ already
looks backward in strain, some early positive histories also measured motion
across the preceding release. These are properties of the sample space rather
than leakage from the future target, yet they prevent a clean distinction
between present state and residual information from the previous event.

The lookback-aware rule addresses both mechanisms. Positive and negative
windows share the onset separation, and every non-affine reference in the
complete input follows trough and recovery. The rule removes only fixed
windows; labels and probabilities remain unchanged. Compact AP remains above
the mean positive-window fraction, whereas its differences from both
loading-history scores become more negative. It also reduces represented-event
coverage to 78.4\% and
69.8\% and shifts the event-weighted median
preceding interval upward under both definitions. Thus the boundary conclusion
does not depend on the original positive-only region, but it applies to a
subset enriched in more widely separated events rather than to every accepted
release. Because the rule was chosen after inspection of the held-out
outcomes, it remains a sensitivity analysis. The original pooled result is
reported separately, and its activity-gated compact-minus-strain interval
includes zero.

\subsection{What the model comparison does and does not imply}

The six compact channels are not an exhaustive material state. They omit
fabric tensors, contact anisotropy, local free volume, soft spots, and
fluidity-like fields. Nonlocal granular-fluidity models explicitly relate flow
state to cooperative rearrangement activity
\citep{henann2014continuum,zhang2017microscopic}, while local structural
descriptors can predict rearrangement propensity in disordered solids
\citep{manning2011vibrational,schoenholz2016structural,
richard2020predicting}. The present result should therefore be read as a bound
on this coarse observable set, not as evidence that material structure is
irrelevant.

The graph comparison is narrower still. The tested graph excludes weak
contacts, mean-pools all particles, and omits explicit stress and pressure. It
also has more parameters than the compact model and changes representation and
classifier together. Its failure to exceed loading history does not constitute
a representation ablation. Weighted all-contact graphs, anisotropic readouts,
localized pooling, or particle-level rearrangement targets remain plausible
alternatives \citep{mandal2022robust,patinoramirez2023percolating}.

A more direct physical target would be a conditional slip probability from a
fixed configuration. Multiple continuations with small controlled
perturbations could estimate
\begin{equation}
 \widehat p_{\mathrm{slip}}(X_t)=\frac{1}{M}\sum_{r=1}^{M}
 \mathbf 1\!\left(\text{continuation $r$ slips within the horizon}\right).
 \label{eq:branch-probability}
\end{equation}
This would separate insufficiency of the observed state from variability of a
single deterministic continuation and provide a meaningful upper bound for a
coarse-grained predictor.

\subsection{Scope and limitations}

The trajectories cover startup shear to total strain 0.50 at one system size,
one shear rate, and one contact/drag parameter set. Removing pre-yield
windows does not create a statistically stationary regime. Absolute strain
may lose much of its value in long steady-state trajectories, so the present
ordering should not be generalized beyond the simulated startup cohort.
Testing that regime would require substantially longer trajectories and a new
analysis specified before their inspection.

The temporal nulls answer narrower questions than a generative physical model.
Circular shifting preserves each compact-score sequence and only changes its
phase on retained endpoints; exceeding it does not establish information
conditional on all loading coordinates. The onset surrogate conditions on
event count and duration while retaining the observed validity geometry; it
does not synthesize a mechanically consistent stress or particle trajectory.
Accordingly, neither null identifies a hazard law or attributes the residual
compact alignment to a particular microscopic mechanism.

The two preparations in each $(\phi,\mu)$ cell limit condition-specific
inference. Exact enumeration removes Monte Carlo error from the stated
resampling design, but it does not increase the 18 independent test units or
adjust the post-test comparisons for multiplicity. The intervals also
condition on the fixed three-seed ensemble rather than propagating
training-seed variation. Condition-level AP varied substantially, but two preparations
per cell cannot resolve how the ordering depends on $(\phi,\mu)$.
Forecasting here means retrospective ranking of admissible windows;
calibration, alarm duration, event size, experimental transfer, and online
false-alarm control were not evaluated. Finally, the
release-localization analysis in Appendix~\ref{app:exploratory-audits} did not
meet its event-coverage target. It indicates that non-affine concentration
develops during matchable releases, not that concentration predicts onset.
\section{Conclusions}
\label{sec:conclusions}

On 18 held-out granular-shear trajectories, a compact state model ranked
near-slip windows above the mean positive-window fraction under two nested
event definitions. On the post-test lookback-aware windows, it also exceeded a
within-trajectory circular-phase control. The compact sequence therefore
contains temporally aligned information about imminent slip rather than an
arbitrary phase of the same score history.

That positive result has two boundaries. First, cumulative strain and causal
prefix-replay elapsed strain ranked the same windows more accurately in the
equal-trajectory analysis. Second, the lookback-aware correction retained only
78.4\% and 69.8\% of
the originally represented events and selected longer preceding intervals.
A count-conditioned onset surrogate placed final-catalog age AP inside the
protocol-defined geometry-null ranges under both definitions. High elapsed
ranking therefore cannot by itself identify a renewal mechanism.

These analyses are descriptive post-test sensitivities, not new confirmation.
They show that the tested state descriptors provide protocol-dependent,
temporally aligned retrospective discrimination, while not isolating an
incremental state-specific precursor beyond loading history during startup
shear. Isolating such a precursor will require stationary loading protocols,
richer structural observables, or branching continuations from matched
configurations.
\section*{Statements and Declarations}

\subsection*{Data and code availability}

The compact data and analysis code supporting the reported tables and figures
are archived in the cited repository \cite{zhou2026granular_data}. The record
contains the frozen prediction arrays, event and quality summaries,
configurations, endpoint-eligibility masks, per-trajectory statistics,
exact-resampling and temporal-null results, analysis protocols, final figure
assets, and code needed to reproduce the reported tabulations from those
arrays. The larger particle-resolved HDF5 trajectories and model checkpoint
binaries are available from the corresponding author on reasonable request
because of archive size.

\subsection*{Use of AI-assisted tools}

Generative-AI tools assisted with language drafting and code review. The
authors independently verified the source code, calculations, citations,
figures, and interpretation and retain responsibility for the manuscript. No
generative-AI image was used.

\subsection*{Competing interests}

The authors declare no competing interests.

\bibliographystyle{unsrtnat}
\bibliography{references}

\clearpage
\appendix
\section{Supplementary numerical and statistical details}
\label{app:protocol}

\subsection{Cohorts and analysis chronology}

The original 36 trajectories formed a condition-balanced development set: 18
were assigned to training, nine to validation, and nine to the nominal test
subset. During iterative development, that nominal test subset was inspected
while the event audit, controls, and history-matched comparison were refined.
Those results are reported as development-stage estimates.

Before the new trajectories were generated, simulation settings, both event
definitions, original endpoint construction, fitted normalizers, eight
three-seed ensembles, checkpoints, metrics, and condition-stratified sampling
were fixed. The held-out cohort was evaluated without retraining or
normalization refitting. The eight ensembles are the four listed model
pipelines under each of the two event definitions. The separately reported
loading-and-event-history MLP was a development-stage model and was not a ninth
member of this confirmation family. The archived data and code record contains
the prediction scores, configurations, evaluator, eligibility masks, and
numerical summaries needed for the reported analyses.

Two new preparations were generated in each $(\phi,\mu)$ cell. All
18 held-out trajectories passed the pre-existing
numerical criteria, and none was replaced or selected according to event
count, prevalence, AP, or model ordering. Exact trajectory identifiers are
provided in the archived data and code record.

\subsection{Numerical protocol, preprocessing, and quality checks}

Table~\ref{tab:appendix-numerical-protocol} collects numerical details that
support reproduction without interrupting the main physical argument.

\begin{longtable}{@{}>{\raggedright\arraybackslash}p{0.29\textwidth}
  >{\raggedright\arraybackslash}p{0.65\textwidth}@{}}
  \caption{Reproducibility details for simulation, labels, inputs, and
  independent-test evaluation.}\label{tab:appendix-numerical-protocol}\\
  \toprule
  Quantity & Definition or value \\
  \midrule
  \endfirsthead
  \toprule
  Quantity & Definition or value \\
  \midrule
  \endhead
  Preparation / production step & $10^{-4}$ / $4\times10^{-4}$ \\
  Contact-scale regime & Across 26,955,862 stored contacts,
    overlap divided by pair diameter had median 0.011,
    97.5th percentile 0.045, and maximum
    0.141. The range of $k_n$ divided by
    trajectory-median pressure was 38.4--286.9. \\
  Damped contact times & Small--small / small--large / large--large
    restitution: 0.568/0.520/0.447; damped collision duration in
    production steps: 71/82/101. \\
  Loading and drag times & $1/\zeta=0.500$; one saved-frame
    interval $=0.333$; the forecast horizon spans
    10.000 drag times. \\
  Compression and consolidation & Compression from $\phi=0.58$ in 16 stages;
    zero-shear convergence checks every 250 preparation steps; velocities and
    angular velocities multiplied by 0.5 every 500 steps. \\
  Normalized kinetic-energy criterion &
    $(K/N)/(k_n\bar d^2)\leq10^{-7}$, where $\bar d$ is mean diameter. \\
  Force-imbalance criterion &
    $\langle\lVert\mathbf F_i^{\rm net}\rVert\rangle_i/
    \langle f_n\rangle_{\rm contacts}\leq0.05$. \\
  Pressure-stability criterion &
    $(\max |P|-\min |P|)/\operatorname{mean}|P|\leq0.02$ over the latest five
    checks. All three criteria had to pass after at least 0.5 reduced time units. \\
  Observed consolidation & All 54 preparations converged in
    11,750--63,750 preparation steps; independent-test preparations converged in
    12,750--34,750 steps. \\
  Independent-test inertial number &
    $I=|\dot\gamma|\bar d/\sqrt{|P|/\rho}$; the trajectory-level 90th
    percentiles ranged from $3.27\times10^{-4}$ to $9.45\times10^{-4}$. \\
  Independent-test force imbalance & The trajectory-level 90th percentiles of the
    production force-imbalance ratio ranged from $1.54\times10^{-3}$ to
    $4.98\times10^{-3}$. \\
  Independent-test completeness & Every trajectory contained 500 ordered frames,
    reached $\gamma=0.4998$, had effective-rate ratio one to numerical
    precision, and had $\Dmin$-valid fraction 0.982. \\
  $\Dmin$ affine fit & Backward strain lag 0.010 and reference-neighbor cutoff
    2.2 small-particle diameters. The $2\times2$ normal matrix received $\lambda I$, with
    $\lambda=10^{-12}\max[\operatorname{tr}(X^{\mathsf T}X),1]$. The first
    valid saved frame uses the production initial configuration at $\gamma=0$
    as its reference; that configuration is stored in the in-memory history
    but maps to reference-frame index $-1$. Hence 491 of 500 saved frames are
    lag-valid. \\
  Particle-level $\Dmin$ validity & The development set had 17,676 valid-lag
    frames and 9,050,112 particle fits; the independent test set had 8,838
    valid-lag frames and 4,525,056 fits. No fit in either set received the
    invalid-fit zero sentinel, and no non-finite value occurred. \\
  Yield fallback & If no accepted prominent peak existed, the threshold was
    90\% of the median smoothed stress over the final 30\% of frames. Yield was
    the first two-frame persistent crossing after $\gamma=0.005$. The fallback
    was unused in all 54 trajectories. \\
  Label smoothing & Seven-frame centered moving average for yield and
    five-frame centered moving average for post-yield slip; both used reflected
    boundaries and were confined to offline label construction. \\
  Stress--pressure ratio &
    $\sigma_{xy}/\max(|P|,\epsilon)$ with
    $\epsilon=\max[10^{-12},10^{-8}\operatorname{median}(|P_{\rm train}|)]
    =7.158481\times10^{-7}$. \\
  Compact normalization & Per-channel median and IQR fitted on complete
    development-training trajectories only and loaded unchanged for independent testing. \\
  Graph normalization & Full-contact particle coordination used median
    $4.0$ and IQR $2.0$; $D^2_{\mathrm{scale}}=
    1.82695614\times10^{-4}$ was the median finite positive particle value on
    valid-lag frames; radius used mean $0.599999994$ and standard deviation
    $0.099999994$; and $f_{\mathrm{scale}}=140.868683$ was the median
    $|f_n|$ over retained strong-contact edges. All values were fitted on
    complete development-training trajectories and were identical in the
    graph checkpoints across labels and seeds. \\
  Class weights & $w_c=n/(2n_c)$ from development-training windows only. \\
  Training settings & AdamW, learning rate $10^{-3}$, weight decay $10^{-4}$,
    gradient-norm clipping at one, 30 epochs; batch size 8 for compact/control
    MLPs and 4 for graph models. No optimization used independent-test data. \\
  MLP parameter counts & $H=4$ compact: 5,955 instantiated / 5,890
    classification-path parameters; stress--pressure--ratio control: 5,187 /
    5,122; $H=8$ compact: 7,491 / 7,426. The 65-parameter auxiliary yield head
    had zero loss weight. \\
  Graph parameter count & 55,235 instantiated
    parameters, compared with 7,491 for the
    $H=8$ compact model (ratio 7.37). \\
  Original pooled and secondary intervals & 10,000
    condition-stratified draws; two trajectories sampled with replacement
    within each of nine cells, carrying all windows of each selected trajectory
    together. \\
  Corrective mean per-trajectory intervals & Exact probability-weighted
    enumeration of all $3^9=19,683$ within-cell
    compositions under the same two-trajectory resampling design. \\
  \bottomrule
\end{longtable}

\FloatBarrier

\section{Additional held-out evaluation results}
\label{app:independent-details}

\subsection{Event-definition comparison}

The complete onset-count comparison is reported in
Table~\ref{tab:appendix-event-overlap}.

\begin{table}[htbp]
  \centering
  \caption{Independent-test event counts and onset overlap. ``Within
  tolerance'' uses the same $\Delta\gamma=0.010$ separation used to merge
  nearby candidates.}
  \label{tab:appendix-event-overlap}
  \begin{tabular}{@{}lr@{}}
    \toprule
    Quantity & Count \\
    \midrule
    Activity-gated onsets & 88\\
    Activity-ungated onsets & 106\\
    Additional activity-ungated onsets & 18\\
    Activity-gated onsets matched exactly & 88\\
    Activity-gated onsets matched within tolerance & 88\\
    Trajectories with changed event list & 11\\
    \bottomrule
  \end{tabular}
\end{table}

For context, the development set contained
182 activity-gated and 209
activity-ungated onsets; 181 activity-gated onsets matched
exactly and all
182 matched within tolerance. The independent test set
reproduces this event-overlap pattern without using event count as a retention
criterion.

\subsection{Forecast-window sets}

Table~\ref{tab:appendix-window-sets} reports the admissible observations under
each label definition and input history.

\begin{table}[htbp]
  \centering
  \caption{Independent-test window, positive-window, and distinct-event counts.
  Differences between $H=4$ and $H=8$ reflect complete-history availability;
  differences between definitions additionally reflect their event lists and
  exclusion regions.}
  \label{tab:appendix-window-sets}
  \begin{tabular}{@{}llrrr@{}}
    \toprule
    Definition & History & Windows & Positive windows & Events \\
    \midrule
    Activity-gated & $H=4$ & 6166&
      1230& 88\\
    Activity-gated & $H=8$ & 5856&
      1163& 84\\
    Activity-ungated & $H=4$ & 5945&
      1396& 106\\
    Activity-ungated & $H=8$ & 5586&
      1287& 97\\
    \bottomrule
  \end{tabular}
\end{table}

The finite-tail rule preserves accepted events. The event column counts events
that contribute at least one positive observation time at the specified history.

\subsection{Seed-level and ensemble results}

Table~\ref{tab:appendix-seed-results} separates optimization-seed variation
from the probability ensemble used for the main estimates.

\begin{table}[htbp]
  \centering
  \caption{Fixed seed-level AP, probability-ensemble AP, and ensemble
  prevalence on the independent test set.}
  \label{tab:appendix-seed-results}
  \scriptsize
  \resizebox{\textwidth}{!}{\begin{tabular}{@{}llrrrrr@{}}
    \toprule
    Definition & Model & Seed 3407 & Seed 7349 & Seed 9109 & Ensemble AP & Prevalence \\
    \midrule
    Activity-gated & Compact MLP, $H=4$ & 0.2812&
      0.2792& 0.2585&
      0.284& 0.199\\
    Activity-gated & Stress--pressure--ratio MLP, $H=4$ &
      0.2725& 0.2420&
      0.2662& 0.259&
      0.199\\
    Activity-gated & Compact MLP, $H=8$ & 0.2107&
      0.2386& 0.2420&
      0.234& 0.199\\
    Activity-gated & Graph GNN--LSTM, $H=8$ & 0.2489&
      0.2451& 0.2204&
      0.244& 0.199\\
    Activity-ungated & Compact MLP, $H=4$ & 0.3029&
      0.2805& 0.3138&
      0.305& 0.235\\
    Activity-ungated & Stress--pressure--ratio MLP, $H=4$ &
      0.3242& 0.3361&
      0.3248& 0.337&
      0.235\\
    Activity-ungated & Compact MLP, $H=8$ & 0.2916&
      0.2722& 0.2834&
      0.285& 0.230\\
    Activity-ungated & Graph GNN--LSTM, $H=8$ & 0.2836&
      0.2479& 0.2553&
      0.276& 0.230\\
    \bottomrule
  \end{tabular}}
\end{table}

Ensemble AP is computed after averaging seed probabilities and need not equal
mean seed AP. Seed variation describes optimization variability; the
trajectory bootstrap describes preparation-sampling uncertainty. They are not
combined into one interval.

\subsection{Exact-observation-time paired comparisons}

Table~\ref{tab:appendix-paired-results} gives the complete paired estimates.

\begin{table}[htbp]
  \centering
  \caption{Independent-test paired AP differences. The final column is the
  descriptive fraction of valid condition-stratified bootstrap draws above
  zero.}
  \label{tab:appendix-paired-results}
  \small
  \resizebox{\textwidth}{!}{\begin{tabular}{@{}lrrrl@{}}
    \toprule
    Difference & Common windows & Estimate & 95\% interval & Fraction $>0$ \\
    \midrule
    Activity-gated compact $H=4$ $-$ control & 6166&
      0.025& [-0.016, 0.072]&
      0.873\\
    Activity-ungated compact $H=4$ $-$ control & 5945&
      -0.032& [-0.071, 0.016]&
      0.085\\
    Activity-gated compact $H=8$ $-$ graph & 5856&
      -0.010& [-0.046, 0.028]&
      0.295\\
    Activity-ungated compact $H=8$ $-$ graph & 5586&
      0.009& [-0.013, 0.031]&
      0.762\\
    \bottomrule
  \end{tabular}
  }
\end{table}

Exact $(\text{trajectory},t_{\rm end})$ keys were aligned before resampling, and
the same within-condition trajectory draw was applied to both score sets.
Intervals that include zero do not establish equivalence.

\clearpage

\section{Sensitivity analyses}
\label{app:exploratory-audits}

Unless stated otherwise, results in this section were specified after the
original held-out summary had been inspected and are reported as
sensitivities.

\subsection{Endpoint eligibility and pooled-window diagnostics}

In trajectory $j$, the original loader converted its fixed strain buffer to
\begin{equation}
 b_j=\left\lceil
 \frac{0.020}{\widetilde{\Delta\gamma}_j}-10^{-6}
 \right\rceil
 \label{eq:appendix-buffer-conversion}
\end{equation}
saved-frame intervals, where $\widetilde{\Delta\gamma}_j$ is the median saved
strain increment. With $\widetilde{\Delta\gamma}_j=9.996\times10^{-4}$, this
gave $b_j=21$ in every held-out trajectory.

The original four-frame endpoint set allowed a positive history to re-enter
nine frames after a previous onset, while the minimum observed elapsed offset
among negative endpoints was 25 frames under both event definitions. The
positive-only region contained 231 and
336 windows. The corresponding
recency score was $q_t=-(t-s^{-}(t))$ when a previous accepted onset existed
and $q_t=-10^9$ otherwise; equal offsets were tied. Its AP values fell from
0.434 and 0.523 to
0.194 and
0.227 after those windows were removed.
The diagnostic ranks recency in the retrospectively accepted event catalog and
is reported only as a post-test check of endpoint geometry.

Table~\ref{tab:appendix-pooled-eligibility} gives pooled point estimates under
all three window rules. The trajectory-prevalence diagnostic assigns a
constant score within each trajectory equal to that trajectory's held-out
positive-window prevalence. It therefore measures the susceptibility of pooled
AP to between-trajectory prevalence variation and must not be interpreted as a
fitted baseline.

\begin{table}[htbp]
  \centering
  \caption{Pooled-window AP under the window sensitivities. Elapsed strain is
  reconstructed by causal prefix replay. The positive-fraction diagnostic uses
  evaluation labels and is shown only as a nuisance check.}
  \label{tab:appendix-pooled-eligibility}
  \footnotesize
  \setlength{\tabcolsep}{3.2pt}
  \begin{tabular}{@{}llrrrrr@{}}
    \toprule
    Definition & Rule & Positive frac. & Compact & Strain & Causal elapsed & Fraction diag. \\
    \midrule
    Activity-gated & Original & 0.199&
      0.284& 0.331&
      0.329& 0.306\\
    Activity-gated & Common exclusion & 0.168&
      0.248& 0.287&
      0.286& 0.252\\
    Activity-gated & Lookback-aware & 0.164&
      0.245& 0.290&
      0.290& 0.246\\
    Activity-ungated & Original & 0.235&
      0.305& 0.364&
      0.344& 0.375\\
    Activity-ungated & Common exclusion & 0.189&
      0.245& 0.296&
      0.284& 0.301\\
    Activity-ungated & Lookback-aware & 0.185&
      0.241& 0.301&
      0.286& 0.297\\
    \bottomrule
  \end{tabular}
\end{table}

The complete-history $D^2_{\min}$ support reached to or before the previous
onset for 81 activity-gated and
131 activity-ungated positive windows.
The lookback-aware mask combines the common onset separation with the
requirement that the earliest stored reference frame follow the previous
trough plus recovery. Per-endpoint masks, previous onset/trough indices,
reference indices, per-trajectory AP rows, and their hashes are provided in the
archived data and code record.

\subsection{Window-geometry and temporal-null audit}

Table~\ref{tab:appendix-event-selection} reports event rather than window
coverage. An accepted event contributes once if at least one positive endpoint
targets its onset. The interval columns are event weighted and omit first
events. They therefore quantify which releases remain represented after the
lookback-aware correction.

\begin{table}[htbp]
  \centering
  \caption{Event selection by the lookback-aware window rule. Coverage is
  relative to events represented by the original positive windows. Preceding
  intervals are onset-to-onset saved-frame intervals with every retained event
  weighted once.}
  \label{tab:appendix-event-selection}
  \small
  \resizebox{\textwidth}{!}{\begin{tabular}{@{}lrrcc@{}}
    \toprule
    Definition & Original events & Lookback events & Coverage &
    Median preceding interval, original $\rightarrow$ lookback \\
    \midrule
    Activity-gated & 88 &
      69 & 78.4\% &
      48.0 $\rightarrow$
      78.0 \\
    Activity-ungated & 106 &
      74 & 69.8\% &
      39.0 $\rightarrow$
      72.5 \\
    \bottomrule
  \end{tabular}
  }
\end{table}

The fixed-bin positive-window fractions conditional on frames since the
previous onset are also included in the archived data and code record. They are
forecast-window fractions, not instantaneous hazards. Their nonmonotone pattern under
the selected endpoint rules was not used to infer a waiting-time law.

Table~\ref{tab:appendix-temporal-nulls} separates two null questions. The
circular-phase control shifts the frozen compact scores on retained endpoints.
The onset surrogate instead rebuilds endpoints and evaluates an ideal
final-catalog age score. Consequently, its catalog-age column is not the
causal prefix-replay score in the main model comparison.

\begin{table*}[htbp]
  \centering
  \caption{Post-test temporal-null audit. Brackets are central 95\%
  randomization ranges, not trajectory-resampling confidence intervals. The
  phase null uses 100,000 cohort compositions; the hard-core onset
  null uses 2,000 surrogate cohorts.}
  \label{tab:appendix-temporal-nulls}
  \small
  \setlength{\tabcolsep}{4.2pt}
  \resizebox{\textwidth}{!}{\begin{tabular}{@{}lccccc@{}}
    \toprule
    Definition & Compact AP & Circular-phase null & Phase upper tail &
    Catalog-age AP & Hard-core onset null \\
    \midrule
    Activity-gated & 0.310 &
      0.211 [0.171, 0.257] &
      0.00005 & 0.527 &
      0.461 [0.383, 0.538] \\
    Activity-ungated & 0.325 &
      0.239 [0.196, 0.287] &
      0.00051 & 0.504 &
      0.503 [0.428, 0.572] \\
    \bottomrule
  \end{tabular}
  }
\end{table*}

The compact score exceeded the phase-null 97.5th percentile under both event
definitions. Both actual catalog-age AP values remained inside the onset-null
central ranges; their one-sided upper-tail probabilities were
0.048 and 0.479. The first is
near the upper tail and is not evidence of equivalence. Neither result tests a
Poisson process, memorylessness, or a specific renewal law.

\subsection{Temporal activity aggregation replay}

The stored activity-gated label takes the maximum of the framewise mean
activity over the 10\% most active particles from onset through trough,
inclusive. A
full detector replay
reproduced all 88 stored onsets exactly. Replacing only the
temporal aggregation by the inclusive mean produced
71 events; additionally excluding the trough produced
68; and the inclusive median produced
62. Thus the lower event count does not follow from
``temporal mean'' alone. None of these replayed lists replaced the frozen
events or entered model selection.

\subsection{Dependence of label variants and development-cohort reversal}

The activity-gated and activity-ungated primary statistics share the same
trajectories and resampling draws. Their AP-excess correlation was
0.874 in 10,000 shared draws and
0.875 under probability weighting of all
19,683 compositions. The two label results therefore
probe detector robustness but do not constitute independent replication.

Table~\ref{tab:appendix-development-time} reports the development-test
quantities that motivated explicit discussion of cohort reversal. This subset
had already been viewed during model development and is descriptive.

\begin{table}[htbp]
  \centering
  \caption{Development-test ordering and nested-stacker increments. The
  time-plus-condition stacker used one validation trajectory per condition for
  fitting.}
  \label{tab:appendix-development-time}
  \small
  \resizebox{\textwidth}{!}{\begin{tabular}{@{}lrrrrr@{}}
    \toprule
    Definition & Compact AP & Strain AP & Compact $-$ strain & Time increment & Time+condition increment \\
    \midrule
    Activity-gated & 0.405&
      0.296&
      0.109&
      0.029&
      0.012\\
    Activity-ungated & 0.476&
      0.326&
      0.150&
      0.051&
      0.029\\
    \bottomrule
  \end{tabular}
  }
\end{table}

\subsection{Loading-progress, post-yield, and threshold results}

Table~\ref{tab:appendix-time-audit} reports both the prevalence-referenced and
the directly paired strain-referenced contrasts.

\begin{table}[htbp]
  \centering
  \caption{The all-frozen compact-minus-prevalence entries reproduce the
  original prespecified primary checks. Cumulative-strain comparisons and all
  post-yield entries were specified after the independent outcomes were
  inspected. Intervals use 10,000 condition-stratified trajectory draws.}
  \label{tab:appendix-time-audit}
  \small
  \resizebox{\textwidth}{!}{\begin{tabular}{@{}llrrrrll@{}}
    \toprule
    Definition & Window set & Windows & Compact AP & Prev. & $\gamma_{\rm obs}$ AP & Compact $-$ prev. (95\% interval) & Compact $-$ $\gamma_{\rm obs}$ (95\% interval) \\
    \midrule
    Activity-gated & Original & 6166& 0.284&
      0.199& 0.331&
      0.085 [0.038, 0.152]&
      -0.047 [-0.101, 0.013]\\
    Activity-gated & Post-yield & 4492& 0.314&
      0.258& 0.338&
      0.055 [0.018, 0.102]&
      -0.025 [-0.065, 0.015]\\
    Activity-ungated & Original & 5945& 0.305&
      0.235& 0.364&
      0.070 [0.035, 0.130]&
      -0.059 [-0.111, -0.004]\\
    Activity-ungated & Post-yield & 4285& 0.338&
      0.310& 0.371&
      0.028 [-0.007, 0.082]&
      -0.033 [-0.079, 0.010]\\
    \bottomrule
  \end{tabular}}
\end{table}

At validation-selected thresholds 0.069824 and
0.073074, all-window balanced accuracy was
0.599 [0.563, 0.637] and
0.599 [0.568, 0.631]. Without recalibration, post-yield
balanced accuracy was 0.517 [0.491, 0.543] and
0.510 [0.471, 0.548].

\begin{figure*}[htbp]
  \centering
  \includegraphics[width=\textwidth]{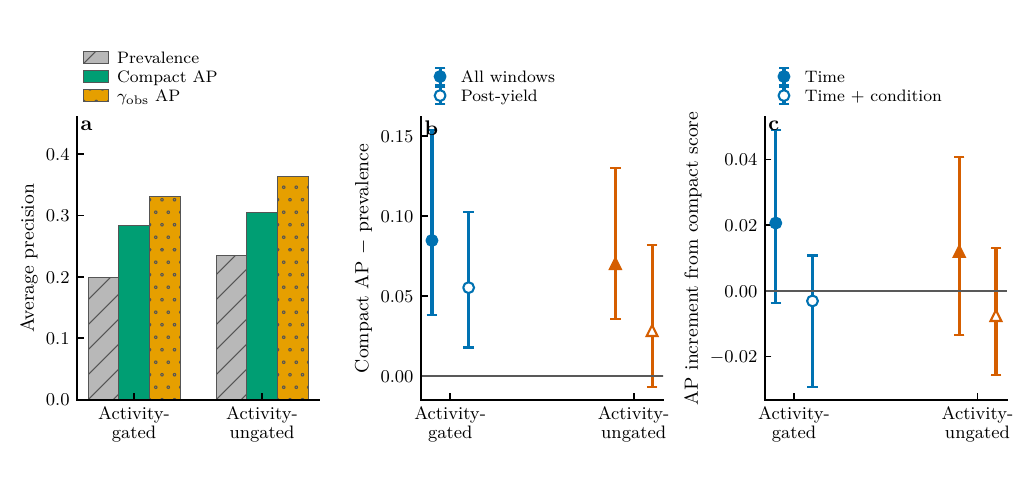}
  \caption{Loading-progress audit. (a)~Prevalence, compact-model AP, and
  zero-parameter cumulative-strain AP on the original $H=4$ endpoints.
  (b)~Compact AP minus prevalence on all endpoints and on the post-yield
  subset. The all-endpoint compact-minus-prevalence checks were prespecified;
  cumulative-strain and post-yield comparisons were introduced after the
  frozen summary was inspected. (c)~AP increment obtained by adding the fixed
  compact score to development-fitted loading-time stackers, without and with
  condition indicators. Error bars are percentile 95\% intervals from
  10,000 condition-stratified trajectory draws. Panel (c)
  is post-test}
  \label{fig:exploratory-time-audit}
\end{figure*}

Adding the fixed compact score to a development-fitted loading-time stacker
changed AP by 0.021 [-0.004, 0.049]
for activity-gated labels and 0.012
[-0.014, 0.041] for activity-ungated labels. After condition
indicators were added, the corresponding increments were
-0.003 [-0.029, 0.011] and
-0.008 [-0.026, 0.013].
Every interval included zero. Figure~\ref{fig:exploratory-time-audit}
summarizes these post-test checks.

\FloatBarrier

\subsection{Elapsed-strain baselines and exact resampling enumeration}

The main-text elapsed-strain score replays the complete fixed detector on every
trajectory prefix. Table~\ref{tab:appendix-causal-recurrence} summarizes its
lookback-aware and pre-first-event checks. All full-trajectory replays exactly
reproduced the stored yield, onset, and trough arrays.

\begin{table}[htbp]
  \centering
  \caption{Post-test causal-availability audit on lookback-aware windows.
  ``Post-first'' removes endpoints before the prefix replay has detected its
  first event. Brackets are exact 95\% intervals from the finite
  condition-stratified resampling design.}
  \label{tab:appendix-causal-recurrence}
  \small
  \begin{tabular}{@{}llrl@{}}
    \toprule
    Definition & Endpoint subset & Causal elapsed AP & Compact $-$ causal elapsed (95\% interval) \\
    \midrule
    Activity-gated & Lookback-aware &
      0.534 &
      -0.223
      [-0.288, -0.159] \\
    Activity-ungated & Lookback-aware &
      0.508 &
      -0.184
      [-0.224, -0.143] \\
    Activity-gated & Post-first &
      0.652 &
      -0.332
      [-0.402, -0.261] \\
    Activity-ungated & Post-first &
      0.593 &
      -0.251
      [-0.298, -0.204] \\
    \bottomrule
  \end{tabular}
\end{table}

An earlier post-test audit instead exposed a stored onset after its
final-catalog trough plus two recovery frames. That retrospective score is retained
in Table~\ref{tab:appendix-recurrence} to document the analysis chronology; it
is not the causal elapsed score used in the main comparison. The
development-trained loading-and-event-history MLP exists only for the
activity-gated definition.

\begin{table}[htbp]
  \centering
  \caption{Earlier retrospective temporal baselines on the independent set.
  Every paired difference is computed on exact common endpoints; intervals use
  10,000 condition-stratified trajectory draws.}
  \label{tab:appendix-recurrence}
  \small
  \begin{tabular}{@{}llrl@{}}
    \toprule
    Definition & Score & AP & Compact AP minus score AP (95\% interval) \\
    \midrule
    Activity-gated & Final-catalog elapsed score & 0.281&
      0.003 [-0.057, 0.068]\\
    Activity-ungated & Final-catalog elapsed score & 0.295&
      0.010 [-0.033, 0.067]\\
    Activity-gated & \makecell[l]{Loading-and-event-\\history MLP} & 0.326&
      -0.042 [-0.112, 0.032]\\
    \bottomrule
  \end{tabular}
\end{table}

All 19,683 compositions of the nine-cell resampling design
were also evaluated. Table~\ref{tab:appendix-exact-enumeration} compares the
Monte Carlo percentile interval with the probability-weighted central interval
over the exact finite distribution.

\begin{table}[htbp]
  \centering
  \caption{Loading-progress comparisons under Monte Carlo bootstrap and exact
  enumeration. Negative values favor cumulative strain.}
  \label{tab:appendix-exact-enumeration}
  \small
  \resizebox{\textwidth}{!}{\begin{tabular}{@{}llll@{}}
    \toprule
    Definition / endpoints & Paired difference & Bootstrap estimate (95\% interval) & Exact central 95\% interval \\
    \midrule
    Activity-gated / $H=4$ & Compact $-$ strain &
      -0.047 [-0.101, 0.013]&
      [-0.100, 0.012]\\
    Activity-ungated / $H=4$ & Compact $-$ strain &
      -0.059 [-0.111, -0.004]&
      [-0.112, -0.004]\\
    Activity-gated / $H=8$ & Compact $-$ strain &
      -0.100 [-0.160, -0.045]&
      [-0.161, -0.044]\\
    Activity-ungated / $H=8$ & Compact $-$ strain &
      -0.075 [-0.142, -0.011]&
      [-0.143, -0.013]\\
    Activity-gated / $H=8$ & Graph $-$ strain &
      -0.090 [-0.150, -0.029]&
      [-0.152, -0.028]\\
    Activity-ungated / $H=8$ & Graph $-$ strain &
      -0.083 [-0.146, -0.024]&
      [-0.146, -0.025]\\
    \bottomrule
  \end{tabular}}
\end{table}

\subsection{Event-scale and spatial audit}

Table~\ref{tab:appendix-event-physics} summarizes the independent-cohort event
scales and the event--quiet spatial comparison. Quiet values use every
admissible label-negative $H=4$ endpoint, and the paired interval is computed
after averaging each class within trajectory.

\begin{table}[htbp]
  \centering
  \caption{Exploratory event and non-affine concentration statistics. A
  participation ratio $q_{D^2}=1$ denotes uniform activity, with smaller values
  denoting stronger concentration. Event and quiet columns are pooled medians;
  the final column is a paired mean difference formed after within-trajectory
  averaging and therefore is not their arithmetic difference.}
  \label{tab:appendix-event-physics}
  \small
  \resizebox{\textwidth}{!}{\begin{tabular}{@{}lrrrrrrl@{}}
    \toprule
    Definition & Events & Median drop/$P$ & Median duration $\Delta\gamma$ & Median interevent $\Delta\gamma$ & Event $q_{D^2}$ & Quiet $q_{D^2}$ & Event $-$ quiet (95\% interval) \\
    \midrule
    Activity-gated & 88& 0.028&
      0.010& 0.048&
      0.108& 0.178&
      -0.073 [-0.086, -0.060]\\
    Activity-ungated & 106& 0.025&
      0.009& 0.039&
      0.108& 0.179&
      -0.073 [-0.084, -0.061]\\
    \bottomrule
  \end{tabular}}
\end{table}

The corresponding median fractions of total activity carried by the most
active 10\% of particles were
0.704/0.605
at activity-gated event/quiet frames and
0.703/0.603
under the activity-ungated definition. Event frames are the activity maxima
during the onset-to-trough release intervals, selected from the trace based on
the 10\% most active particles. This is not a pre-onset forecast.

The original comparison applies a temporal maximum only to the event interval.
Table~\ref{tab:appendix-matched-localization} reports the duration- and
selector-matched quiet-window control. Matching was performed within
trajectory, with all quiet frames required to be admissible label-negative
endpoints and the quiet start required to lie within
$|\Delta\gamma|\leq0.035$ of onset. Quiet starts were not
reused. The matching protocol and thresholds were fixed before the
concentration values were computed.

\begin{table}[htbp]
  \centering
  \caption{Exploratory duration- and selector-matched quiet-window control.
  Differences are
  event minus matched quiet values; brackets are condition-stratified
  trajectory-bootstrap 95\% intervals. Before the post-test spatial contrasts
  were computed, the adequacy target was set to at least 80\% event coverage
  with all 18 trajectories represented.}
  \label{tab:appendix-matched-localization}
  \small
  \resizebox{\textwidth}{!}{\begin{tabular}{@{}lrrrrll@{}}
    \toprule
    Definition & Events & Matched & Coverage & Trajectories & Activity-maximum $q_{D^2}$ difference & Onset $q_{D^2}$ difference \\
    \midrule
    Activity-gated & 88& 60&
      68.2\%& 18 &
      -0.044 [-0.056, -0.031]&
      0.040 [0.018, 0.062]\\
    Activity-ungated & 106& 64&
      60.4\%& 18 &
      -0.042 [-0.053, -0.030]&
      0.026 [0.003, 0.049]\\
    \bottomrule
  \end{tabular}}
\end{table}

The matched activity-maximum top-10\% share differences were
0.069
[0.053, 0.084] and
0.069
[0.052, 0.084]. All 18 trajectories contributed,
but only 68.2\% and 60.4\% of events
had a complete same-trajectory quiet interval within the fixed audit caliper.
The control therefore supports a matchable-event sensitivity, not a catalog-wide
estimate. The positive onset participation differences and negative
activity-maximum differences place the selected concentration change within
the release.

\begin{figure*}[htbp]
  \centering
  \includegraphics[width=\textwidth]{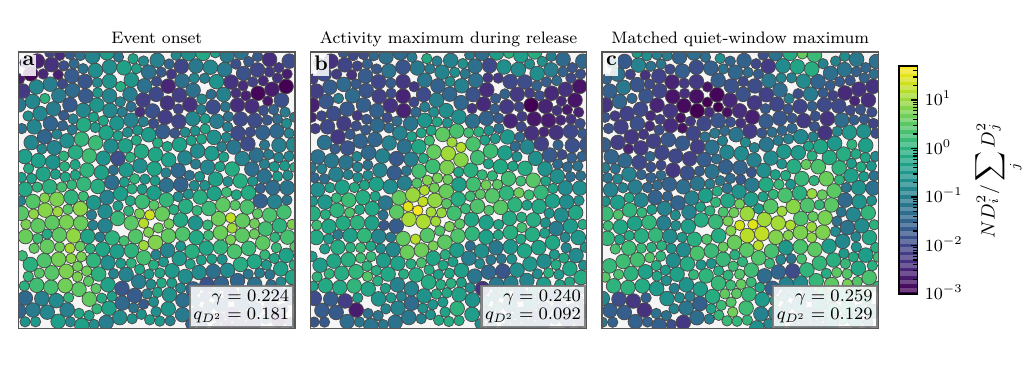}
  \caption{Illustrative matched pair under the activity-gated definition at
  $\phi=0.85$ and $\mu=0.3$. The pair was selected before map inspection as
  the activity-gated match whose event-minus-quiet participation difference
  lay closest to the median. (a)~Event onset. (b)~Within-release activity
  maximum. (c)~Matched quiet-window activity maximum. Colors show
  $ND_i^2/\sum_jD_j^2$ on one logarithmic scale. Smaller $q_{D^2}$ denotes
  stronger concentration. The maps illustrate the exploratory matched subset
  and do not enter its estimates}
  \label{fig:representative-particle-map}
\end{figure*}

The particle maps in Fig.~\ref{fig:representative-particle-map} were chosen
only for illustration. Before any map was inspected, the fixed rule calculated
the participation ratio at the within-release activity maximum minus that at
the matched quiet-window maximum for all 60 activity-gated pairs. It selected
the pair whose difference was closest to the median; trajectory name and event
index resolved any exact tie. The selection record, input hashes, and shared
color scale are supplied in the archived data and code record. This presentation
choice does not enter any estimate or interval.

At $\phi=0.84/0.85/0.86$, the fractions in the near-threshold band
$0.010\leq\Delta\sigma/P<0.015$ were
21.6\%/7.4\%/29.2\% for activity-gated events and
18.2\%/15.2\%/31.0\% for activity-ungated events. No
alternative drop threshold was used for model evaluation.

The slip search begins strictly after yield. A separate replay showed that the
yield-following drop exceeded the largest accepted activity-gated drop in only
16 of 54 trajectories and did not in
38. Thus post-yield detection structurally
excludes the yield peak but does not systematically exclude each trajectory's
largest drop.

Among endpoints in the 15 frames immediately before yield, the activity-gated set
contained 264 windows
(69 positive and
195 negative), while the
activity-ungated set contained 250
(69 positive and
181 negative). In the retained
post-yield subset, 58
activity-gated and 54
activity-ungated four-frame
histories still included the yield frame; the wider backward-$\Dmin$ support
could cross yield for 169 and
165 endpoints. These counts
are why the analysis is described as a post-yield endpoint sensitivity rather
than a fully post-yield state-history estimand.

\subsection{Resolution of pipeline comparisons}

The pooled-window AP difference and the equal-trajectory mean answer different
questions. The MDE calculation belongs to the latter estimand and is presented
beside it in Table~\ref{tab:appendix-mde-estimands}; it is not used to relabel a
pooled comparison as equivalent.

\begin{table}[htbp]
  \centering
  \caption{Estimand audit for descriptive two-sided MDE calculations
  ($\alpha=0.05$, 80\% power, $n=18$ trajectory differences).}
  \label{tab:appendix-mde-estimands}
  \small
  \resizebox{\textwidth}{!}{\begin{tabular}{@{}lrrr@{}}
    \toprule
    Comparison & Pooled-window AP difference & Equal-trajectory mean $\pm$ SD & MDE \\
    \midrule
    Activity-gated compact $H=4$ $-$ control & 0.025&
      0.036 $\pm$ 0.171&
      0.120\\
    Activity-ungated compact $H=4$ $-$ control & -0.032&
      -0.009 $\pm$ 0.149&
      0.105\\
    Activity-gated compact $H=8$ $-$ graph & -0.010&
      0.017 $\pm$ 0.183&
      0.128\\
    Activity-ungated compact $H=8$ $-$ graph & 0.009&
      0.061 $\pm$ 0.108&
      0.076\\
    \bottomrule
  \end{tabular}
  }
\end{table}

\begin{figure*}[htbp]
  \centering
  \includegraphics[width=\textwidth]{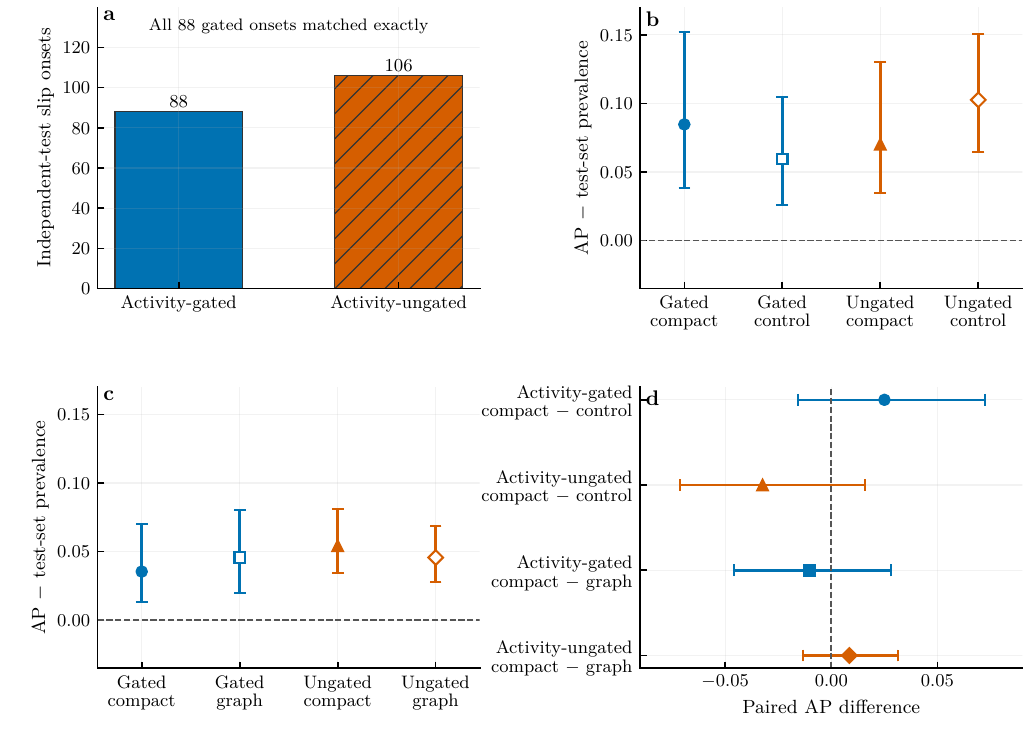}
  \caption{Original held-out pipeline comparisons retained as secondary
  analyses. (a)~Event counts and exact onset overlap. (b)~Pooled AP minus
  prevalence for the four-frame compact and stress--pressure--ratio models.
  (c)~Pooled AP minus prevalence for the history-matched eight-frame compact
  and graph models. (d)~Observation-time-matched AP differences. Error bars
  are 95\% intervals from condition-stratified trajectory resampling. These
  pooled-window comparisons use the original endpoint rule and are not the
  lookback-aware mean per-trajectory analysis in the main text}
  \label{fig:secondary-pipelines}
\end{figure*}

\subsection{Graph training history}

The graph network has 55,235 parameters compared with
7,491 in the history-matched compact model.
Table~\ref{tab:appendix-graph-training} reports the saved histories; no
additional retraining was performed.

\begin{table}[htbp]
  \centering
  \caption{Saved graph-model training histories. Validation AP selected the
  checkpoint; ``loss'' gives the first and final training loss over 30 epochs.}
  \label{tab:appendix-graph-training}
  \small
  \begin{tabular}{@{}lrrrrr@{}}
    \toprule
    Definition & Seed & Best epoch & Best validation AP & Final validation AP & Loss, first $\to$ final \\
    \midrule
    Activity-gated & 3407 & 27 & 0.381 & 0.372 & 0.655 $\to$ 0.074 \\
    Activity-gated & 7349 & 19 & 0.358 & 0.339 & 0.662 $\to$ 0.057 \\
    Activity-gated & 9109 & 22 & 0.393 & 0.377 & 0.656 $\to$ 0.056 \\
    Activity-ungated & 3407 & 10 & 0.445 & 0.345 & 0.659 $\to$ 0.055 \\
    Activity-ungated & 7349 & 13 & 0.382 & 0.315 & 0.660 $\to$ 0.054 \\
    Activity-ungated & 9109 & 16 & 0.413 & 0.371 & 0.660 $\to$ 0.052 \\
    \bottomrule
  \end{tabular}
\end{table}

\FloatBarrier

\section{Condition-resolved and development-stage results}
\label{app:secondary-results}

\subsection{Descriptive condition-level results}

Table~\ref{tab:appendix-condition-results} resolves the primary compact-model
results across the nine simulated conditions.

\begin{table}[htbp]
  \centering
  \caption{Condition-level results for the $H=4$ compact ensemble. Each row
  contains two trajectories. These values are descriptive; no subgroup
  inference is made.}
  \label{tab:appendix-condition-results}
  \scriptsize
  \resizebox{\textwidth}{!}{\begin{tabular}{@{}rrrrrrrrrrrr@{}}
    \toprule
    & & \multicolumn{5}{c}{Activity-gated} & \multicolumn{5}{c}{Activity-ungated} \\
    \cmidrule(lr){3-7}\cmidrule(lr){8-12}
    $\phi$ & $\mu$ & Windows & Events & AP & Prev. & $\gamma$ AP & Windows & Events & AP & Prev. & $\gamma$ AP \\
    \midrule
    0.84 & 0.1 & 594 & 16 & 0.399 & 0.350 & 0.556 & 553 & 20 & 0.427 & 0.438 & 0.568 \\
    0.84 & 0.3 & 662 & 10 & 0.517 & 0.208 & 0.298 & 631 & 12 & 0.385 & 0.242 & 0.302 \\
    0.84 & 0.5 & 648 & 11 & 0.314 & 0.235 & 0.425 & 639 & 12 & 0.308 & 0.252 & 0.463 \\
    0.85 & 0.1 & 643 & 12 & 0.398 & 0.261 & 0.425 & 609 & 17 & 0.595 & 0.343 & 0.641 \\
    0.85 & 0.3 & 714 & 9 & 0.298 & 0.172 & 0.389 & 683 & 10 & 0.395 & 0.202 & 0.377 \\
    0.85 & 0.5 & 723 & 6 & 0.142 & 0.124 & 0.175 & 723 & 6 & 0.123 & 0.124 & 0.175 \\
    0.86 & 0.1 & 662 & 11 & 0.367 & 0.240 & 0.306 & 635 & 14 & 0.388 & 0.285 & 0.327 \\
    0.86 & 0.3 & 791 & 7 & 0.140 & 0.129 & 0.406 & 760 & 8 & 0.151 & 0.154 & 0.381 \\
    0.86 & 0.5 & 729 & 6 & 0.205 & 0.123 & 0.239 & 712 & 7 & 0.210 & 0.147 & 0.298 \\
    \bottomrule
  \end{tabular}}
\end{table}

\FloatBarrier

\subsection{Interpretation boundaries}

The event-definition comparison removes only the $\Dmin$ activity criterion;
other label choices were not tested. The graph and compact pipelines differ in
both representation and architecture, and the control comparison is not a test
of the physical relevance of excluded variables. Event concentration is
measured at the within-release activity maximum and does not establish a
pre-onset spatial precursor. The duration- and selector-matched quiet-window
control covered only
68.2\% and 60.4\% of the two
catalogs, so it supports only a matchable-event sensitivity. The 18
trajectories, not their thousands of windows, are the independent test units;
exact enumeration changes numerical integration over the resampling design,
not that sample size. AP measures retrospective ranking; calibration,
prospective false-alarm policy, stiffness and drag sensitivity, finite-size
scaling, rate transfer, and experimental transfer remain outside the analysis.

\subsection{Development estimates retained for transparency}
\label{app:development-results}

The 36-trajectory development set informed construction of the analysis, so its
nominal test subset is retained only as exploratory context.
Table~\ref{tab:appendix-development-comparison} compares all four paired
estimates with the independent test set; the development-stage
activity-ungated compact--graph ordering was not reproduced.

\begin{figure*}[htbp]
  \centering
  \includegraphics[width=\textwidth]{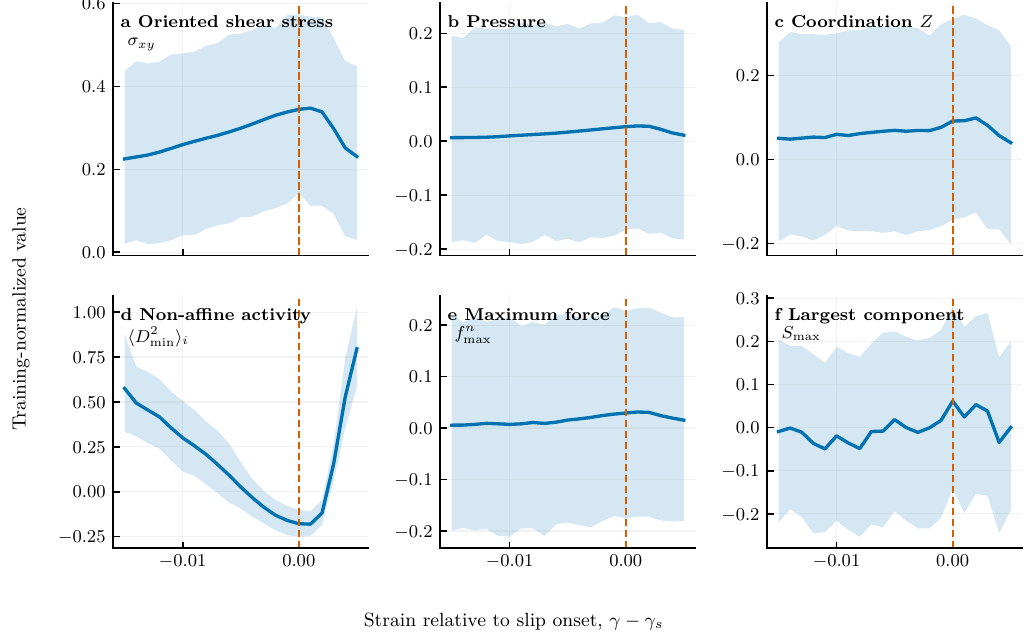}
  \caption{Event-aligned compact observables in the development set: 162
  cleanly alignable activity-gated slips out of 182 from 36 trajectories. Each
  event has a complete $[-15,+5]$ frame window, is more than 20 frames after
  the preceding onset, and more than five frames before the next onset. Values
  are normalized with development-training medians and interquartile ranges.
  Events are averaged first within trajectory, and bands show 95\%
  trajectory-bootstrap intervals. The dashed line denotes onset. Panel (a)
  shows oriented shear stress, so an accepted onset is a local maximum.
  $\langle\Dmin\rangle_i$ is the all-particle mean, $f^n_{\max}$ is evaluated over all
  contacts, and $S_{\max}$ is the largest-component particle count at
  $\alpha_f=1.2$. The activity-gated event selection conditions the
  post-onset non-affine profile; it is not evidence of a pre-onset precursor.
  These profiles are descriptive and do not enter the independent tests}
  \label{fig:event-aligned}
\end{figure*}

\begin{table}[htbp]
  \centering
  \caption{Selected development-stage paired AP differences versus their
  independent-test estimates. Development intervals used nine
  trajectory clusters and are exploratory.}
  \label{tab:appendix-development-comparison}
  \small
  \resizebox{\textwidth}{!}{\begin{tabular}{@{}lll@{}}
    \toprule
    Difference & Development estimate (95\% interval) & Independent-test estimate (95\% interval) \\
    \midrule
    Activity-gated compact $H=4$ $-$ control &
      0.020 [-0.064, 0.119]&
      0.025 [-0.016, 0.072]\\
    Activity-ungated compact $H=4$ $-$ control &
      0.080 [-0.014, 0.154]&
      -0.032 [-0.071, 0.016]\\
    Activity-gated compact $H=8$ $-$ graph &
      0.115 [-0.021, 0.251]&
      -0.010 [-0.046, 0.028]\\
    Activity-ungated compact $H=8$ $-$ graph &
      0.160 [0.015, 0.249]&
      0.009 [-0.013, 0.031]\\
    \bottomrule
  \end{tabular}}
\end{table}

\FloatBarrier

\end{document}